\documentclass{emulateapj}
\begin{document}

\newcommand{\kms}{\ensuremath{\mathrm{km}\,\mathrm{s}^{-1}}}
\newcommand{\etal}{et al.}
\newcommand{\LCDM}{$\Lambda$CDM}
\newcommand{\ML}{\ensuremath{\Upsilon_{\star}}}
\newcommand{\MLmax}{\ensuremath{\Upsilon_{max}}}
\newcommand{\MLpop}{\ensuremath{\Upsilon_{pop}}}
\newcommand{\MLopt}{\ensuremath{\Upsilon_{acc}}}
\newcommand{\Om}{\ensuremath{\Omega_m}}
\newcommand{\OL}{\ensuremath{\Omega_{\Lambda}}}
\newcommand{\A}{\ensuremath{{\cal A}}}
\newcommand{\Q}{\ensuremath{{\cal Q}}}
\newcommand{\Pop}{\ensuremath{{\cal P}}}
\newcommand{\G}{\ensuremath{{\Gamma}}}
\newcommand{\Gopt}{\ensuremath{{\Gamma_{\star}}}}
\newcommand{\Lsun}{\ensuremath{L_{\odot}}}
\newcommand{\Msun}{\ensuremath{{\cal M}_{\odot}}}
\newcommand{\mass}{\ensuremath{{\cal M}}}
\newcommand{\Mst}{\ensuremath{{\cal M}_{\star}}}
\newcommand{\Mg}{\ensuremath{{\cal M}_g}}
\newcommand{\Md}{\ensuremath{{\cal M}_d}}
\newcommand{\Mh}{\ensuremath{{\cal M}_h}}
\newcommand{\Sd}{\ensuremath{{\Sigma}_0}}
\newcommand{\fst}{\ensuremath{f_{\star}}}
\newcommand{\Vst}{\ensuremath{V_{\star}}}
\newcommand{\vst}{\ensuremath{v_{\star}}}
\newcommand{\Vg}{\ensuremath{V_{g}}}
\newcommand{\Vb}{\ensuremath{V_{b}}}
\newcommand{\Vh}{\ensuremath{V_h}}
\newcommand{\Vf}{\ensuremath{V_{f}}}
\newcommand{\gn}{\ensuremath{g_N}}
\newcommand{\csb}{\ensuremath{\mu_0}}
\newcommand{\magsq}{\ensuremath{\mathrm{mag.}\,\mathrm{arcsec}^{-2}}}
\newcommand{\surfdens}{\ensuremath{{\cal M}_{\odot}\,\mathrm{pc}^{-2}}}
\newcommand{\MDAC}{MDAcc}
\newcommand{\HI}{H{\sc i}}

%\shorttitile{The Baryonic Tully-Fisher Relation}
%\shortauthors{McGaugh}

\title{The Baryonic Tully-Fisher Relation of Galaxies with
Extended Rotation Curves \\
and the Stellar Mass of Rotating Galaxies} 

\author{Stacy S.~McGaugh} 

\affil{Department of Astronomy, University of Maryland}
\affil{College Park, MD 20742-2421}    
\email{ssm@astro.umd.edu}

\begin{abstract}
I investigate the Baryonic Tully-Fisher relation for a sample of galaxies
with extended 21 cm rotation curves
spanning the range $20 \lesssim \Vf \le 300\,\kms$.
A variety of scalings of the stellar mass-to-light ratio \ML\ are considered.
For each prescription for \ML, I give fits of the form $\Md = \A \Vf^x$.
Presumably, the prescription that comes closest to the correct value will
minimize the scatter in the relation.  The fit with minimum scatter has 
$\A = 50\, \Msun\, \textrm{km}^{-4}\, \textrm{s}^4$ and $x = 4$.
This relation holds over five decades in mass.
Galaxy color, stellar fraction, and \ML\ are correlated with each
other and with \Md, in the sense that more massive galaxies tend to be
more evolved.  There is a systematic dependence of the degree of maximality 
of disks on surface brightness.  High surface brightness galaxies typically
have $\ML \sim \threequarters$ of the maximum disk value, 
while low surface brightness galaxies typically attain $\sim \onequarter$
of this amount.
\end{abstract}


\section{Introduction}

The Tully-Fisher relation (Tully \& Fisher 1977) has played an important role
in establishing the extragalactic distance scale (e.g., Pierce \& Tully 1988;
Sakai \etal\ 2000).  In this context, it has been treated as a simple and convenient 
empirical relation between luminosity and line-width.  
The reason why it works is also important, particularly to our understanding
of galaxies as physical objects and how they formed
(e.g., Eisentstein \& Loeb 1996; McGaugh \& de Blok 1998a; 
Courteau \& Rix 1999; van den Bosch 2000; Navarro \& Steinmetz 2000a,b).
Now that the issue of the distance scale is widely considered to be settled,
we can hope to place an absolute scale on galaxy mass as well as luminosity.

The physical basis of the Tully-Fisher relation is widely presumed to be a
relation between a galaxy's total mass and rotation velocity (e.g., Freeman 1999).
Luminosity is a proxy, being proportional to stellar mass, which in turn
depends on the total mass.  
McGaugh \etal\ (2000) found that a more fundamental relationship
between the baryonic mass and rotation velocity does indeed exist, 
provided that both stellar and gas mass are considered
(Milgrom \& Braun 1998).

The relation resulting from the sum of stellar and gas mass is referred to as the 
Baryonic Tully-Fisher (BTF) relation.  The BTF has also been investigated by
Bell \& de Jong (2001), Verheijen (2001), Gurovich \etal\ (2004), and 
Pfenniger \& Revaz (2005).  These efforts find broadly similar results, 
in that there is such a relation.  However, details of the relation differ.
McGaugh \etal\ (2000) found a steep slope ($x \approx 4$) from a sample
dominated by galaxies with $H$ or $I$-band photometry and rotation velocities
estimated from the line-width $W_{20}$.  Verheijen (2001) found much the 
same from $K'$-band photometry and flat rotation velocities measured from
resolved 21 cm cubes.  Both these authors assumed a constant value of the
stellar mass-to-light ratio for all galaxies.
Using Verheijen's data and a grid of stellar population
models to refine the estimate of stellar mass,
Bell \& de Jong (2001) found a somewhat shallower slope ($x \approx 3.5$).
Gurovich \etal\ (2004) find a break in the relation, with
low mass galaxies following a steeper slope.

There is an equal variety in the physical interpretations.  
McGaugh \etal\ (2000) argue that the regularity of the BTF implies that, 
after the observed stars and gas are accounted for, no further comparably 
massive reservoirs of baryons are likely to exist in disk galaxies.  
Pfenniger \& Revaz (2005) use the same data to argue the opposite case:
the scatter is somewhat reduced if there are dark baryons weighing several
times the observed gas mass.  In the context of \LCDM, one would expect
still more dark baryons in order to match the universal baryon fraction 
(e.g., Mo \& Mao 2004), though these need not be associated with the disk.

The situation at present remains confused.
The purpose of this paper is to provide the best empirical BTF relation
possible with the currently available data.
Much depends on the value of the stellar mass. 
I consider the effects on the BTF of varying the stellar mass over
a broad range.  Mass-to-light ratios are scaled
by several recipes:  as a fraction of maximum disk; with respect to stellar 
population synthesis models; and by the Mass Discrepancy---Acceleration 
(\MDAC) relation (McGaugh 2004).  
This treats MOND (Milgrom 1983) as a purely phenomenological prescription. 
A grid of BTF fits are given that covers 
essentially any plausible choice of stellar mass-to-light ratio.  The scatter of 
the relation varies with this choice, and an optimal choice that minimizes
the scatter in the BTF relation is clear.

\begin{deluxetable*}{lccccccccc}
\tabletypesize{\small}
\tablewidth{0pt}
\tablecaption{Galaxy Data\label{RCdata}}
\tablehead{
\colhead{Galaxy} & \colhead{\Vf} & \colhead{\Mst} & \colhead{\Mg} &
 \colhead{\csb} & \colhead{$R_d$} & \colhead{$B-V$} & \colhead{\MLmax} &
 \colhead{\MLpop} & \colhead{\MLopt} \\
 \colhead{} & \colhead{(\kms)} & \multicolumn{2}{c}{($10^{10}\;\Msun$)} & \colhead{} &
 \colhead{(kpc)} & \colhead{} & \multicolumn{3}{c}{($\Msun/\Lsun$)} }
 \startdata
  UGC 2885  	&300 	&30.8\phn\phn  & 5.0\phn\phn  &22.0 &13.0  &[0.47] & \phn 2.6 & 0.8 &1.5  \\
  NGC 2841  	&287 	&32.3\phn\phn  & 1.7\phn\phn  &21.1	&\phn 4.6 &0.74	& \phn 7.9 &2.2	&3.8  \\
  NGC 5533  	&250 	&19.0\phn\phn  & 3.0\phn\phn  &23.0	&11.4 	&0.77	&27.0	&2.5 &3.4  \\
  NGC 6674  	&242 	&18.0\phn\phn  & 3.9\phn\phn  &22.5	&\phn 8.3 &0.57	&18.0	&1.1 &2.6  \\
  NGC 3992  	&242 	&15.3\phn\phn  & 0.92\phn &20.4	&\phn 4.1 &0.72	&\phn 5.8	&2.1 &4.9  \\
  NGC 7331  	&232 	&13.3\phn\phn  & 1.1\phn\phn &21.5	&\phn 4.5 &0.63	&\phn 5.8 &1.4	&2.5  \\
  NGC 3953  	&223 	&\phn7.9\phn\phn  & 0.27\phn &20.6	&\phn 3.9 &0.71	&\phn 2.7 &2.0	&2.7  \\
  NGC 5907  	&214 	&\phn9.7\phn\phn  & 1.1\phn\phn &20.7  &\phn 4.0 &0.78 &\phn 4.7 &2.6	&3.9  \\
  NGC 2998  	&213 	&\phn8.3\phn\phn  & 3.0\phn\phn  &20.3	&\phn 5.4 &0.45	&\phn 2.8 &0.7 &1.2  \\
  NGC \phn 801 	&208 	&10.0\phn\phn  & 2.9\phn\phn  &21.9	&12.0 	&0.61	&\phn 8.4 &1.3 &1.4  \\
  NGC 5371  	&208 	&11.5\phn\phn  & 1.0\phn\phn  &21.4  &\phn 7.9 &0.65 &\phn 3.4 &1.6 &1.6  \\
  NGC 5033  	&195 	&\phn 8.8\phn\phn  & 0.93\phn &23.0	&\phn 5.8 &0.55	&\phn 4.6 &1.0 &4.6  \\
  NGC 3893  	&188 	&\phn 4.20\phn & 0.56\phn &20.3 &\phn 2.3 &[0.56] &\phn 2.6 & 1.1 &2.0  \\
  NGC 4157  	&185 	&\phn 4.83\phn & 0.79\phn &21.3	&\phn 5.0 &0.66	&\phn 3.1 &1.6 &2.4  \\
  NGC 2903  	&185 	&\phn 5.5\phn\phn  & 0.31\phn &20.5	&\phn 2.0 &0.55	&\phn 3.6 &1.0 &3.6  \\
  NGC 4217  	&178 	&\phn 4.25\phn & 0.25\phn &21.4	&\phn 4.2 &0.77	&\phn 2.4 &2.5 &2.2  \\
  NGC 4013  	&177 	&\phn 4.55\phn & 0.29\phn &21.1	&\phn 3.5 &0.83	&\phn 4.0 &3.2 &3.1  \\
  NGC 3521  	&175 	&\phn 6.5\phn\phn  & 0.63\phn &20.5  &\phn 2.4 &0.68 &\phn 3.8 &1.8 &2.7  \\
  NGC 4088  	&173 	&\phn 3.30\phn & 0.79\phn &20.9	&\phn 3.1 &0.51	&\phn 1.7 &0.9 &1.1  \\
  NGC 3877  	&167 	&\phn 3.35\phn & 0.14\phn &20.1	&\phn 3.0 &0.68	&\phn 2.4 &1.8 &1.7  \\
  NGC 4100  	&164 	&\phn 4.32\phn & 0.30\phn &20.1	&\phn 2.8 &0.63	&\phn 2.9 &1.4 &2.4  \\
  NGC 3949  	&164 	&\phn 1.39\phn & 0.33\phn &19.6	&\phn 1.7 &0.39	&\phn 1.0 &0.6 &0.8  \\
  NGC 3726  	&162 	&\phn 2.62\phn & 0.62\phn &21.6	&\phn 5.5 &0.45	&\phn 1.8 &0.7 &1.0  \\
  NGC 6946  	&160 	&\phn 2.7\phn\phn & 2.7\phn\phn &21.7 &\phn 5.6 &0.40 &\phn 1.1 &0.6 &0.5  \\
  NGC 4051  	&159 	&\phn 3.03\phn & 0.26\phn &21.2	&\phn 4.2 &0.62	&\phn 1.9 &1.4 &1.2  \\
  NGC 3198  	&156 	&\phn 2.3\phn\phn & 0.63\phn &21.6 &\phn 2.6 &0.43 &\phn 3.1 &0.6 &2.6  \\
  NGC 2683  	&155 	&\phn 3.5\phn\phn & 0.05\phn &21.1  &\phn 1.2 &0.65 &\phn 7.0 &1.6 &5.8  \\
  NGC 3917  	&135 	&\phn 1.4\phn\phn & 0.18\phn &22.1 &\phn 2.9 &0.60 &\phn 2.4 &1.3 &1.3  \\
  NGC 4085  	&134 	&\phn 1.0\phn\phn & 0.13\phn &20.0 &\phn 1.6 &0.47 &\phn 1.4 &0.8 &1.2  \\
  NGC 2403  	&134 	&\phn 1.1\phn\phn & 0.47\phn &21.4 &\phn 2.1 &0.39 &\phn 1.5 &0.6 &1.4  \\
%\tablebreak
  NGC 3972  	&134 	&\phn 1.0\phn\phn & 0.12\phn &20.6 &\phn 2.1 &0.55 &\phn 2.1 &1.0 &1.5  \\
  UGC \phn 128 	&131 	&\phn 0.57\phn & 0.91\phn &24.2	&\phn 9.2 &0.60	&\phn 5.0 &1.3 &1.1  \\
  NGC 4010  	&128 	&\phn 0.86\phn & 0.27\phn &22.5	&\phn 2.9 &[0.54] &\phn 2.1 & 1.0 &1.4  \\
  F568--V1   	&124 	&\phn 0.66\phn & 0.34\phn &23.3	&\phn 3.2 &0.57	&\phn 5.4 &1.1 &3.0  \\
  NGC 3769  	&122 	&\phn 0.80\phn & 0.53\phn &21.1	&\phn 1.7 &[0.64] &\phn 1.4 & 1.5 &1.2  \\
  NGC 6503  	&121 	&\phn 0.83\phn & 0.24\phn &21.9	&\phn 1.7 &0.57	&\phn 1.7 &1.1 &1.7  \\
  F568--3    	&120 	&\phn 0.44\phn & 0.39\phn &23.1	&\phn 4.0 &0.61	&\phn 4.3 &1.3 &1.3  \\
  NGC 4183  	&112 	&\phn 0.59\phn & 0.34\phn &22.8	&\phn 3.4 &0.39	&\phn 1.8 &0.6 &0.7  \\
  F563--V2   	&111 	&\phn 0.55\phn & 0.32\phn &22.1	&\phn 2.1 &0.51	&\phn 3.6 &0.9 &1.8  \\
  F563--1    	&111 	&\phn 0.40\phn & 0.39\phn &23.6	&\phn 4.3 &0.64	&\phn 9.0 &1.5 &3.0  \\
  NGC 1003  	&110 	&\phn 0.30\phn & 0.82\phn &21.6 &\phn 1.9 &0.55 &\phn 0.5 &1.0 &0.2  \\
  UGC 6917  	&110 	&\phn 0.54\phn & 0.20\phn &22.9	&\phn 3.4 &[0.53] &\phn 3.5 & 1.0 &1.4  \\
  UGC 6930  	&110 	&\phn 0.42\phn & 0.31\phn &22.3	&\phn 3.0 &[0.59] &\phn 2.1 & 1.2 &0.8  \\
    M 33    	&107 	&\phn 0.48\phn & 0.13\phn &21.1 &\phn 1.7 &0.55 &\phn 1.2 &1.0 &0.6  \\
  UGC 6983  	&107 	&\phn 0.57\phn & 0.29\phn &23.0 &\phn 3.6 &[0.45] &\phn 4.3 & 0.7 &1.7  \\
  NGC \phn 247 	&107 	&\phn 0.40\phn & 0.13\phn &23.4	&\phn 2.9 &0.54	&\phn 3.0 &1.0 &1.1  \\
  NGC 7793  	&100 	&\phn 0.41\phn & 0.10\phn &20.3 &\phn 1.1 &0.63 &\phn 1.9 &1.4 &1.2 \\
  NGC \phn 300  & \phn 90  &\phn 0.22\phn & 0.13\phn &22.2 &\phn 2.1 &0.58 &\phn 1.4 &1.2 &0.7  \\
  NGC 5585  	& \phn 90  &\phn 0.12\phn & 0.25\phn &21.9 &\phn 1.4 &0.46 &\phn 0.8 &0.7 &0.5  \\
  NGC \phn\phn 55 & \phn 86 &\phn 0.10\phn & 0.13\phn &21.5 &\phn 1.6 &0.54 &\phn 1.0 &1.0 &0.2  \\
  UGC 6667  	& \phn 86  &\phn 0.25\phn & 0.08\phn &23.8 &\phn 2.8 &[0.65] &\phn 2.5 & 1.5 &1.0 \\
  UGC 2259  	& \phn 86  &\phn 0.22\phn & 0.05\phn &22.3 &\phn 1.3 &\nodata &\phn 3.9 &\nodata &2.1 \\
  UGC 6446  	& \phn 82  &\phn 0.12\phn & 0.30\phn &23.1  &\phn 3.1&[0.39] &\phn 2.2 & 0.5 &0.5 \\
  UGC 6818  	& \phn 73  &\phn 0.04\phn & 0.10\phn &23.5 &\phn 1.9 &[0.43] &\phn 0.7 & 0.6 &0.2 \\
  NGC 1560  	& \phn 72  &\phn 0.034 & 0.098	&23.2	&\phn 1.3 &0.57	&\phn 4.1 &1.1 &1.0  \\
  IC 2574   	& \phn 66  &\phn 0.010 & 0.067	&23.4  &\phn 2.2 &0.42 &\phn 0.6  &0.6 &0.1  \\
  DDO 170   	& \phn 64  &\phn 0.024 & 0.061	&24.1	&\phn 1.3 &\nodata &15.0 &\nodata &1.5  \\
  NGC 3109  	& \phn 62  &\phn 0.005 & 0.068 &23.1 &\phn 1.6 &[0.47]  &\phn 1.0 & 0.8 &0.1  \\
  DDO 154   	& \phn 56  &\phn 0.004 & 0.045 &23.2 &\phn 0.5 &0.32 &\phn 2.0  &0.4 &0.1  \\
  DDO 168   	& \phn 54  &\phn 0.005 & 0.032 &23.4 &\phn 0.9 &0.32 &\phn 0.6  &0.4 &0.2  \\
\enddata
\end{deluxetable*}

\section{The Data}

McGaugh \etal\ (2000) have already described the BTF over a large range
of velocity and mass.  This large dynamic range is critical to constraining
the slope of the relation, and also its absolute normalization since low mass
galaxies are frequently gas rich and insensitive to assumptions about \ML.
Few other samples cover such a large dynamic range (e.g., Gurovich \etal\ 
2004).

One thing that stands to be improved is the accuracy of the data.
The data used here are from the sample of Sanders \& McGaugh (2002),
as trimmed for accuracy by McGaugh (2004).  The reader is referred to these
papers, and references therein, for a further description of the sample.
These galaxies all have extended 21 cm maps from which the flat
rotation velocity \Vf\ is measured.  The use of \Vf\ 
provides a considerable improvement in accuracy over line-width
measurements (Verheijen 2001).  The result is a much cleaner BTF relation
that is not afflicted by possible corrections to line-widths for turbulent motion.
Turbulent corrections have the potential to affect the slope of the BTF
by systematically adjusting the line-width inferred rotation velocities
of low mass galaxies.
In addition to resolved atomic gas measurements, all galaxies have detailed
surface photometry from which stellar masses can be estimated.  

The sample is identical to that in McGaugh (2004), with one exception.
After re-examining the data, NGC 2915 has been replaced by UGC 6818.
Though the individual data points for NGC 2915 are quite precise, this galaxy does
not have a well defined \Vf, the quantity of interest for the BTF.  The relative distance
uncertainty for NGC 2915 is also uncomfortably large (Meurer, Mackie, \& 
Carignan 1994; Karachentsev \etal\ 2003).  In contrast, UGC 6818 only
barely missed the cut imposed in McGaugh (2004), and has more robust
global measurements.  Having done this exercise, I can see where further
improvements could be made for individual galaxies, but these are generally
very minor.  These are all the galaxies that are available with an 
obtainably high standard of accuracy.

The data are given in Table 1.  Column 1 gives the name of the galaxy.
Column 2 gives \Vf\ in \kms.  These are the fit values given by Sanders \& McGaugh
(2002) which are the average of the outer points.  
As a test, I have remeasured \Vf\ by eye for the sub-sample of
Verheijen \& Sancisi (2001).  These agree to within a few \kms\ with the values
given by Verheijen (2001) and with those tabulated here.  \Vf\ is easily and
robustly measured, provided only that the rotation curve is extended enough.
Column 3 gives the stellar mass for the mass-to-light ratio
from column 10, and column 4 gives
the gas mass (both in units of $10^{10}\, \Msun$).  Column 5 gives the central
surface brightness of the disk in $B$ \magsq, and column six the exponential
disk scale length.  The $B$-band is the only band-pass in common to all galaxies;
consistent results are found in the subset with $K'$-band data 
(Sanders \& Verheijen 1998; Verheijen 2001; McGaugh 2004).  
Column 7 gives the $B-V$ color obtained from the original source given in
Sanders \& McGaugh (2002) if available, or from 
NED\footnote{This research has made use of the NASA/IPAC Extragalactic Database (NED) which is operated by the Jet Propulsion Laboratory, California Institute of Technology, under contract with the National Aeronautics and Space Administration.}
if not.  Colors measured with CCDs are given preference; if these are not
available, RC3 values are used.  $B-V$ colors in square brackets are inferred 
from other measured colors through stellar population models.  
It was most often the case that
$B-R$ had been measured instead of $B-V$.  The colors are used to infer
the mass-to-light ratio of the stellar population from said models.  Several
possible mass-to-light ratios are given in the last three columns:  maximum
disk in column 8, stellar population synthesis in column 9, and 
\MDAC\ in column 10 (McGaugh 2004).

\begin{figure*}
\epsscale{0.8}
\plotone{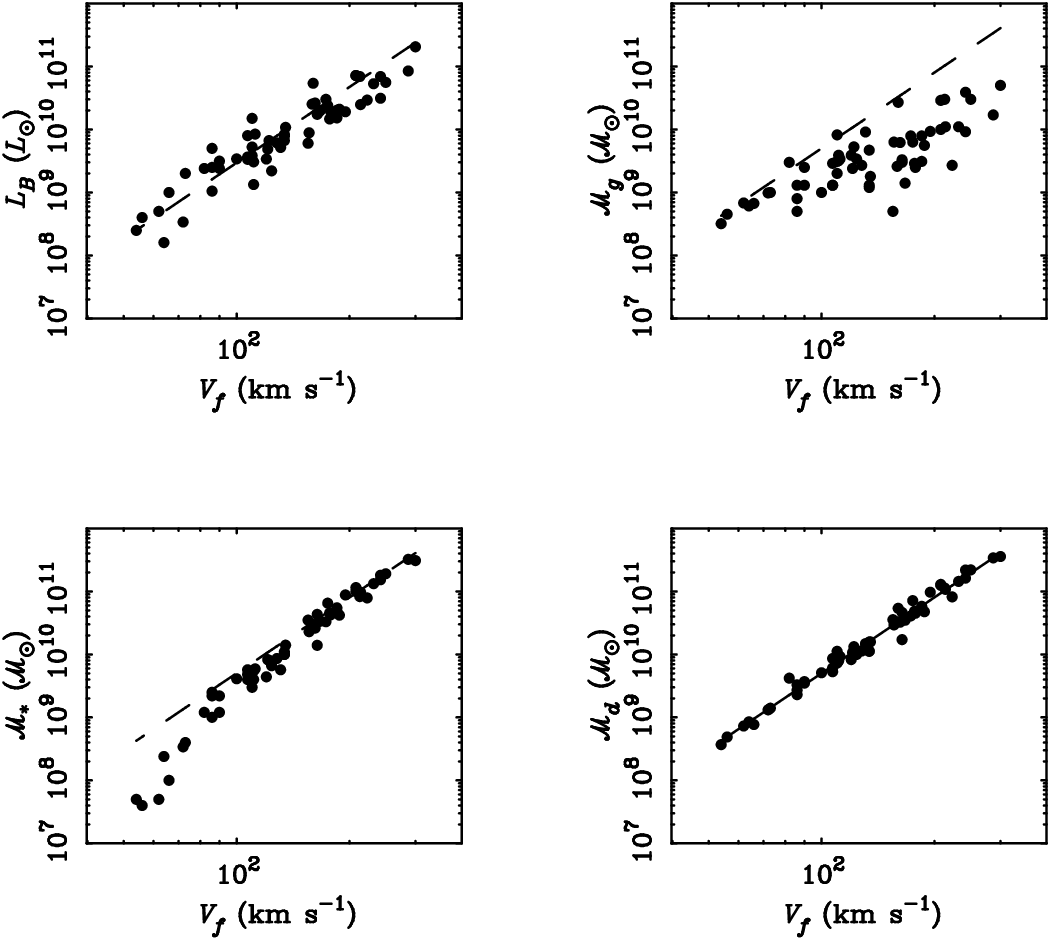}
\caption{Four versions of the Tully-Fisher relation.  The top left panel shows the 
$B$-band luminosity as a function of the flat rotation velocity.  The top right
panel plots the gas mass instead of luminosity.  The bottom left panel plots
stellar mass, $\mass_{\star} = \ML L$ for the \MDAC\ mass-to-light ratios.  
The bottom right panel plots the Baryonic Tully-Fisher relation, with
$\mass_d = \mass_{\star} + \mass_g$.  The solid line is a fit to the data,
$\mass_d = 50 \Vf^4$.  This is drawn as a dashed line in the other panels
for comparison.  In the top left panel, the line is drawn for the mean 
value $\langle \ML \rangle$ $ = 1.7\; \Msun/\Lsun$.
\label{TFintro}}
\end{figure*}

\section{Method}

The BTF is expressed as
\begin{equation}
\Md = \A \Vf^x,
\end{equation}
where \A\ is the normalization and $x$ the slope.  
Note that since mass is used here rather than magnitudes, the slope $m$ in
traditional magnitude units would be $m = -2.5 x$.  \Vf\ is the measured
rotation velocity in the flat part of the rotation curve, 
and \Md\ represents all measured\footnote{In McGaugh \etal\ 
(2000), this was referred to as the baryonic disk mass, hence the subscript.
Both there and here the intent is to represent all known, observed baryons,
regardless of their location in the galaxy (bulge or disk).} baryonic mass.
It includes both stars and gas.  Most of these galaxies are disk dominated,
but bulge mass is included where present (see Sanders 1996).
The gas mass is that measured in neutral hydrogen, corrected for helium and
metals.  Other gas phases are presumed to be negligible.  In effect, the
molecular gas mass is subsumed into the stellar mass (see discussion in
McGaugh 2004).  By the same token, any net internal extinction 
is also subsumed by \ML.  
This can not be a large effect, given the consistency between
$B$ and $K'$ bands (Sanders \& Verheijen 1998).

The baryonic mass of a galaxy is therefor
\begin{equation}
\Md = \Mst +\Mg.
\end{equation}
The stellar mass is given by
\begin{equation}
\Mst = \ML L,
\end{equation}
where \ML\ is the mass-to-light ratio of the stellar population.
Fig.~1 illustrates the various Tully-Fisher relations that can be constructed
from the data in Table 1:  the luminosity-rotation velocity relation, 
$L$-\Vf; this converted into stellar mass, \Mst-\Vf; that with gas only (no stars), 
\Mg-\Vf; and the BTF, \Md-\Vf.

The choice of \ML\ for each galaxy is critical.  
In Fig.~1, the behavior in the \Mst-\Vf\ plane is markedly different from that in
$L$-\Vf.  Use of the \MDAC\ mass-to-light ratio substantially reduces the scatter
when mapping from $L$ to \Mst.  In addition, the break at $\Vf \approx 90\, \kms$
noted by McGaugh \etal\ (2000) becomes apparent in \Mst-\Vf, though it is not
visible in $L$-\Vf.  Constant mass-to-light ratios were used
in McGaugh \etal\ (2000), but the break was apparent there because of the large 
number of very low mass, gas dominated dwarfs.  The data quality restrictions
imposed here exclude those objects.  Simply applying a constant \ML\ to the
galaxies here would preclude the discovery of the break, since this would just be
a shift in the scale of $L$-\Vf.  Yet a specific, well-defined method 
for determining \ML\ recovers the break without input about its existence.

I explore three choices for the stellar mass-to-light ratio:
maximum disk, \MLmax; stellar population synthesis models, \MLpop; 
and that from the \MDAC, \MLopt.
For each of these choices, I construct a grid of \ML\ scaled from each by a 
constant factor: \G, \Pop, or \Q:
%\begin{mathletters}
\begin{eqnarray}
\ML = \G \MLmax \\
\ML = \Pop \MLpop \\
\ML = \Q \MLopt.
\end{eqnarray}
%\end{mathletters}
This provides a prescription for \ML\ that
is specified by method and scaling factor.  Fig.~2 illustrates the BTF 
relations stemming from various choices.  A few details concerning each 
method are worth noting here.

\begin{figure*}
\epsscale{0.8}
\plotone{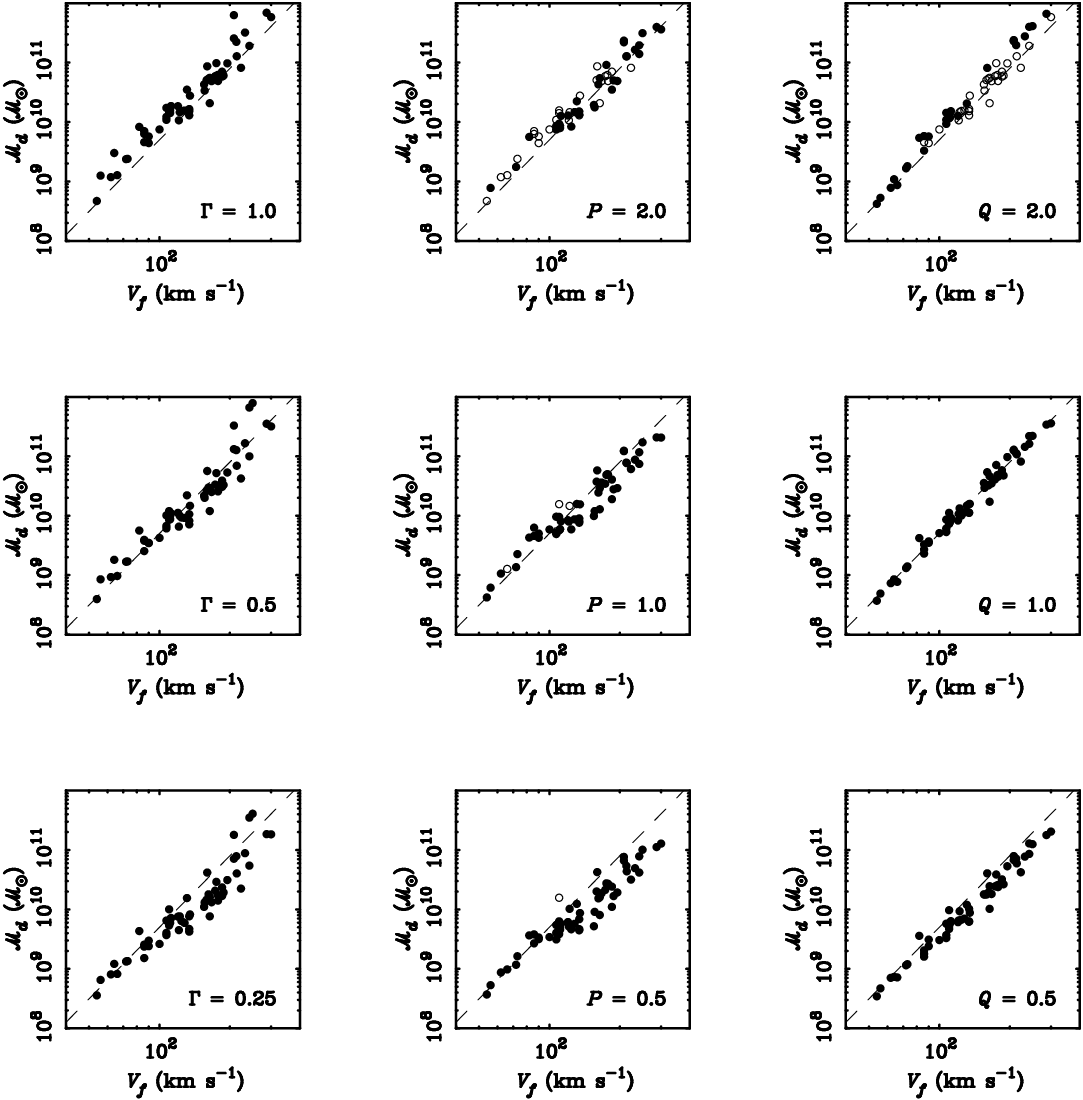}
\caption{The BTF for various choices of stellar mass-to-light ratio.
The left column of panels shows scalings relative to maximum disk:
$\G = 1$, 0.5, and 0.25 from top to bottom.  The middle column shows
scalings relative to the population synthesis models of Bell \etal\ (2003):
$\Pop = 2$, 1, and 0.5 from top to bottom.  Similarly, the right column shows
scalings relative to the mass-to-light ratio from the
\MDAC: $\Q = 2$, 1, 0.5 from top
to bottom.  (Note that $\G = \Pop = \Q = 0$ are all equivalent to the 
gas-only panel in Figure 1.)  The mass-to-light ratio is not allowed
to exceed the maximum disk value.  Galaxies are plotted as open symbols 
with their mass-to-light ratios set to the maximum
disk value if the value specified by \Pop\ or \Q\ 
would have exceeded maximum disk.  Half of the sample has reached this
point by $\Q = 2$.  The fit to the $\Q = 1$ case is shown as a dashed line in
all panels for comparison.
\label{BTFALL}}
\end{figure*}

\subsection{Scaling by Maximum Disk}

Maximum disk is the highest mass-to-light ratio consistent with but not
exceeding the rotation curve data.  The inner shape of rotation curves
are often quite consistent with the shape predicted by the observed
baryons (van Albada \& Sancisi 1986; Selwood 1999; Palunas \& Williams 2000),
leading some to argue that disks must be nearly maximal.  I consider the
range $0 \le \G \le 1$.  

I adopt the maximum disk value given by the original source for the data for
each galaxy (see Sanders \& McGaugh 2002).  Caution should be exercised
in interpreting \MLmax.  Some authors leave room for a dark halo, so that
even ``maximum'' disk may provide only $\sim 84\%$ of the observed velocity
at the peak of the disk contribution (Sackett 1997).  Others leave zero room,
fitting disk-only models as far out as possible (e.g., Palunas \& Williams 2000).
This is a fairly subtle distinction in rotation curve decompositions, but does make
a noticeable difference in \ML.  For example, the average $I$-band mass-to-light
ratio of the Palunas \& Williams (2000) sample is 
$\langle \MLmax \rangle = 2.4\, \Msun/\Lsun$.  
Since $\mass \propto V^2$, if we scaled
this down from a 100\% to 84\% contribution, the mean would be 
$\langle \MLmax \rangle = 1.7\, \Msun/\Lsun$
 (see also Barnes, Sellwood, \& Kosowsky 2004).

In general, there is no uniform definition of maximum disk.  The data originate
from a wide variety of sources, so there is no guarantee as to how maximal
maximum disk is.  Moreover, things can go both ways.  For low surface 
brightness disks, very small differences in the rotation curve can lead to large 
changes in the inferred value of \MLmax\ (Swaters \etal\ 2000; 
McGaugh, Rubin, \& de Blok 2001).  For these galaxies and
some of the more extreme dwarf galaxies, I have re-assessed the value of 
maximum disk.  In spite of these caveats, the original value appears sensible
for most of the galaxies in Table 1.

There are a few exceptions.  For the Ursa Major data (Verheijen \& Sancisi 2001), 
which comprises
a substantial plurality of the data in Table 1, a ``soft'' limit was imposed on the
value of \MLmax\ so that \Vf\ of the fitted pseudo-isothermal halo did not diverge
to absurdly large values (Verheijen 1997).  
This can easily happen for galaxies that are
well described by maximum disk and for which the data are not very extended in
radius.  There is little clear need for a halo in such cases (Palunas \& Williams 2000),
so \Vf\ is poorly constrained and tends towards large values since
only the rising portion of the contribution of the dark matter is seen.  
Unfortunately, imposing this ``soft'' constraint can
sometimes lead to \MLmax\ which is very sub-maximal.  This is a subtle point
which only became apparent because $\MLmax < \MLopt$ 
for four of the galaxies in Table 1 (Sanders \& Verheijen 1998).  This is a 
mathematical impossibility, so I have set $\MLmax = \MLopt$ in these cases.

\subsection{Scaling from Stellar Population Synthesis Models}

Stellar population models have advanced to the point where they give plausible
estimates of the mass-to-light ratio, even for the composite stellar populations of
spiral galaxies.  They are not yet perfect of course, but do provide a decent
choice for estimating \ML\ (Bell \& de Jong 2001; Portinari \etal\ 2004).
Here I employ the models of Bell \etal\ (2003) to estimate the stellar mass-to-light
ratio from the observed color:
\begin{equation}
\log \MLpop = 1.737 (B-V) -0.942
\end{equation}
(from their Table 7).  

The $B-V$ color is given precedence in estimating \MLpop.
When $B-V$ is not available, whatever color is available is used.
The most common substitute is $B-R$, which is expected to be
nearly as well correlated with \ML\ as $B-V$ (Bell \& de Jong 2001).  
The same model (from Table 7 of Bell \etal\ 2003)
is used to estimate $B-V$ (the bracketed colors in Table 1), but
\MLpop\ is based on the observed color.  Credible color information
could not be located for UGC 2259 and DDO 170, the only pieces
missing from Table 1.

\subsection{Scaling from the Mass-Discrepancy---Acceleration Relation}

McGaugh (2004) used the detailed shapes of the rotation curves of the galaxies
in Table 1 to show that there is an empirical relation between acceleration and 
the amplitude of the mass discrepancy (essentially the ratio of dark to
baryonic mass).  This relation holds at every point
along a resolved rotation curve for any non-zero choice of mass-to-light ratio.  
The scatter about this relation depends on this choice;
\MLopt\ is determined by minimizing the scatter with respect to
the mean \textit{local} relation.  
It is interesting to see here how this \ML\ estimator fares with 
the \textit{global} BTF.

The \MDAC\ is a purely empirical relation.  
It is mathematically equivalent to MOND: the \MLopt\ are the same as the
MOND best fit values (Begeman, Broeils, \& Sanders 1991; Sanders 1996;
Sanders \& Verheijen 1998; de Blok \& McGaugh 1998).
Here we must make the distinction between MOND as a fundamental theory
(with its associated difficulties), and as a successful recipe for 
fitting rotation curves.  Only the latter is required.  Indeed, the \MDAC\ is the local
analog of the BTF, and had MOND never been invented, we would perhaps
already have recognized the \MDAC\ purely as an empirical relation (Sancisi 2003).  
Just as the Tully-Fisher relation can be used empirically to estimate distances 
without understanding its physical basis, so too can the \MDAC\ 
be utilized to estimate stellar mass-to-light ratios without prejudice
concerning its theoretical basis.

\section{Results}

Equations (4-6) are used to estimate the stellar mass for the various scalings.
For each scaling, a grid of 10 choices of the scaling constant are made: $\G = 0.1$ 
to 1.0 in steps of 0.1 relative to maximum disk, and for \Pop\ and \Q\ values
ranging from 0.2 to 2.0 in steps of 0.2.  The maximum disk scaling obviously
can not exceed unity, while for the other scalings there is no reason not to consider
values larger than one.  For example, $\Pop > 1$ would simply imply an IMF heavier 
than assumed in the nominal population model which has been adopted.
However, these values should not exceed maximum disk, and are not allowed to
do so.  If the choice of \Pop\ or \Q\ exceeds the maximum disk value for a particular
galaxy, the maximum disk value is used instead.  Half of the sample has
saturated at maximum by $\Q = 2$, so larger values are not considered.

The BTF (logarithm of equation 1) is fit\footnote{These are `direct' fits.
The scatter is small enough that forward and reverse fits are not 
distinguishable except in the limit $\ML \rightarrow 0$.}
to the data in Table 1 for each set 
of choices for \ML.  Every galaxy carries equal weight in the fits.
Since the data have been selected to be of high quality,
the residual uncertainties are likely to be dominated by systematic effects
(such as the precise distance to each galaxy) rather than factors internal
to the data, though these obviously matter as well.
In any case, the choice of mass-to-light ratio completely dominates the results.

\begin{deluxetable}{lccccc}
\tablewidth{0pt}
\tablecaption{BTF Statistics\label{stats}}
\tablehead{
\colhead{} & \colhead{$\log\A$} & \colhead{$\sigma_{\A}$} & \colhead{$x$} &
 \colhead{$\sigma_{x}$} & \colhead{$\sigma_{\mass}$} }
\startdata
$L_B$ & \multicolumn{5}{l}{Luminosity (Normal TF)} \\
1.0 & 2.52       & 0.37   & 3.48   & 0.17   & 0.239 \\ \\
\Mg & \multicolumn{5}{l}{Gas Only} \\
0.0 & 4.67       & 0.48   & 2.28   & 0.48   & 0.313 \\ \\
\G & \multicolumn{5}{l}{Maximum Disk Scaling} \\
0.1 & 3.24       & 0.37   & 3.09   & 0.37   & 0.244 \\
0.2 & 2.76       & 0.35   & 3.38   & 0.35   & 0.230 \\
0.3 & 2.50       & 0.35   & 3.55   & 0.35   & 0.224 \\
0.4 & 2.34       & 0.34   & 3.67   & 0.34   & 0.222 \\
0.5 & 2.23       & 0.34   & 3.75   & 0.34   & 0.220 \\
0.6 & 2.15       & 0.34   & 3.81   & 0.34   & 0.220 \\
0.7 & 2.09       & 0.34   & 3.87   & 0.34   & 0.220 \\
0.8 & 2.05       & 0.34   & 3.91   & 0.34   & 0.220 \\
0.9 & 2.01       & 0.34   & 3.94   & 0.34   & 0.220 \\
1.0 & 1.98       & 0.34   & 3.97   & 0.34   & 0.221 \\ \\
\Pop & \multicolumn{5}{l}{Population Synthesis Scaling} \\  
0.2 & 3.81       & 0.33   & 2.81   & 0.15   & 0.204 \\
0.4 & 3.44       & 0.30   & 3.04   & 0.14   & 0.186 \\
0.6 & 3.23       & 0.29   & 3.19   & 0.14   & 0.183 \\
0.8 & 3.08       & 0.29   & 3.29   & 0.13   & 0.183 \\
1.0 & 2.98       & 0.29   & 3.37   & 0.13   & 0.186 \\
1.2 & 2.87       & 0.28   & 3.44   & 0.13   & 0.188 \\
1.4 & 2.75       & 0.27   & 3.52   & 0.13   & 0.191 \\
1.6 & 2.65       & 0.27   & 3.58   & 0.13   & 0.194 \\
1.8 & 2.54       & 0.26   & 3.64   & 0.12   & 0.197 \\
2.0 & 2.46       & 0.26   & 3.69   & 0.12   & 0.200 \\ \\
\Q & \multicolumn{5}{l}{\MDAC\ Scaling}  \\
0.2 & 3.05       & 0.25   & 3.18   & 0.25   & 0.165 \\
0.4 & 2.46       & 0.20   & 3.53   & 0.20   & 0.129 \\
0.6 & 2.11       & 0.17   & 3.74   & 0.17   & 0.112 \\
0.8 & 1.87       & 0.16   & 3.89   & 0.16   & 0.103 \\
1.0 & 1.70       & 0.15   & 4.00   & 0.15   & 0.098 \\
1.2 & 1.60       & 0.15   & 4.07   & 0.15   & 0.100 \\
1.4 & 1.58       & 0.16   & 4.10   & 0.16   & 0.107 \\
1.6 & 1.60       & 0.18   & 4.10   & 0.18   & 0.118 \\
1.8 & 1.61       & 0.20   & 4.10   & 0.20   & 0.129 \\
2.0 & 1.64       & 0.22   & 4.10   & 0.22   & 0.140 \\
\enddata
\end{deluxetable}

The results of fitting the BTF are given in Table 2.  The intercept $\log\A$ and
slope $x$ are recorded, together with the formal uncertainties in each
($\sigma_{\A}$ and $\sigma_x$).  Also given is the scatter of the data about 
each relation, $\sigma_{\mass}$.  For reference, the input data, in the form
of the normal luminosity-based Tully-Fisher relation is given.  It has a scatter
$\sigma_L = 0.24$, which is equivalent to 0.6 mag.\ (base ten logarithms are used
throughout).  Note also that the limit $\G = \Pop = \Q = 0$ is equivalent to
the gas-only relation, so is given only once.

The results summarized in Table 2 are illustrated in Figs.~3 and 4.  
As the assumed mass-to-light ratio increases, the slope $x$ gradually
increases while the zero point \A\ decreases to compensate.  The lowest
mass galaxies in Table 1 are dominated by gaseous rather than stellar mass,
so the BTF tends to pivot about them.  This can be seen by eye in Fig.~2, 
where the low end of the relation hardly budges while the more massive, 
star dominated galaxies move up and down with \ML.  These low mass,
gas-rich galaxies provide a critical anchor point for the absolute calibration 
of the BTF since their location in this diagram is insensitive to the choice of \ML.

\begin{figure*}
\epsscale{1.0}
\plotone{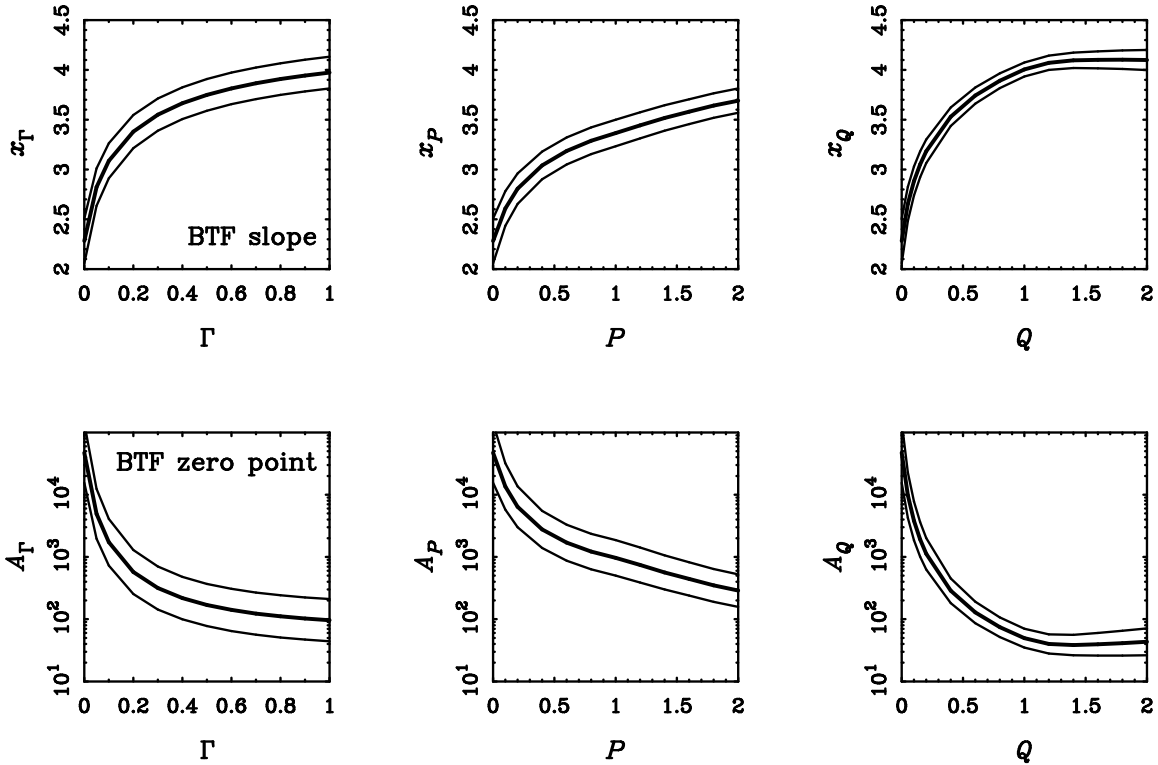}
\caption{The slope (top panels) and zero point (bottom panels) of the BTF for
various scalings of the stellar mass-to-light ratio:  as a fraction of maximum disk 
(left panels), relative to population synthesis models (middle panels), and 
relative to the \MDAC\ mass-to-light ratios (right panels).  
For each choice of \ML, the data have been fit as a straight line
to the logarithm of $\mass_d = \A \Vf^x$ (Table 2).  The thin lines show the formal
uncertainties on the parameters $\log \A$ and $x$ of the fit.  
For all scalings, the slope of the mass---circular
velocity BTF only becomes as shallow as $x = 3$ for implausibly small
mass-to-light ratios.
\label{slopezeropoint}}
\end{figure*}

The scatter $\sigma_{\mass}$ about each fit is shown in Fig.~4.
The maximum disk scaling has scatter comparable to the scatter in the 
input luminosities.  The use of color information with population synthesis
models provides a small increment of improvement in the scatter, as it should
if the models succeed in improving the estimate of \ML\ over a constant value
for all objects.  As stellar mass is rendered unimportant for
very small \G\ and \Pop\ the scatter starts to increase.  
The limit of zero stellar mass, with gas only, obviously makes for an
inadequate BTF, as one would expect.  

\begin{figure}
%\epsscale{0.8}
\plotone{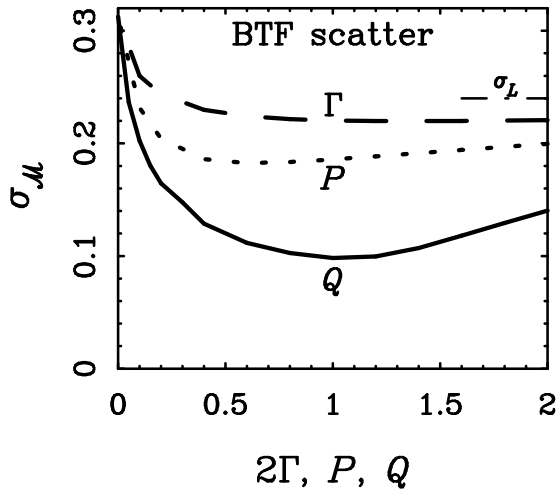}
\caption{The scatter $\sigma_{\mass}$ in the BTF for the various
scalings of mass-to-light ratio.  Dashed line:  maximum disk (plotted against
$2\G$ to fill the same range as the other choices).  Dotted line: population
synthesis.  Solid line: mass-to-light ratios from the \MDAC.  The latter show a 
clear minimum at $\Q = 1$.  The other scalings have no clearly defined minima,
and larger scatter for all choices of \G\ or \Pop.  The ordinary Tully-Fisher 
relation of the input data has a scatter of $\sigma_L = 0.24$ (dash-dot line).
\label{scatterplot}}
\end{figure}

\subsection{The Optimal BTF}

Irrespective of the physics underlying the BTF, we expect that the prescription
for \ML\ that comes closest to the correct value will minimize the scatter about it.
There may be some intrinsic scatter, but if we can get \ML\ right, there should
be no scatter left due to it.  The \MLopt\ prescription gives less scatter than either
\MLmax\ or \MLpop.  There is clear variation in the scatter with \Q, with a well
defined minimum at $\Q = 1$.  
The same mass-to-light ratio chosen to minimize the residuals from the 
local \MDAC\ also minimizes the scatter in the global BTF.
This was already apparent in Fig.~6 of McGaugh (2004), and is
quantitatively confirmed here.  It is rare in extragalactic astronomy that we
find a correlation as strong as the BTF (${\cal R} = 0.99$ for $\Q = 1$).
Indeed, the BTF is so sharply defined that it has a kurtosis of 1.5.

It is difficult to avoid the conclusion that the choice $\Q = 1$ does effectively
give the correct \ML.  This provides an absolute calibration of the BTF:
\begin{equation}
\Md = 50 \Vf^4
\end{equation}
with \Md\ in \Msun\ and \Vf\ in \kms.
Indeed, the scatter in this BTF relation is so small that it could be used
to estimate \ML\ with nearly as great accuracy as the full 
\MDAC, but with considerably less information.
To apply the latter, we require a well resolved rotation curve and \HI\ surface 
density map, and detailed surface photometry.  The BTF requires only global
quantities: $L$, \Mg\ from a single dish 21 cm observation, and an estimate of
\Vf\ (preferably from a rotation curve which is sufficiently resolved to perceive the
flat part, though $\onehalf W_{50}$ would be an adequate substitute, with some
penalty in accuracy.)

This BTF is consistent with the relation first reported by McGaugh \etal\ (2000).
The slope is identical, but the normalization is somewhat larger.  This change is
not particularly significant ($1 \sigma$).  It is largely due to the improvement in 
data quality and the use of \Vf\ rather than $\onehalf W_{20}$.  The line-width 
$W_{20}$ is systematically larger than $2 \Vf$, which is more closely approximated
by $W_{50}$ (Broeils 1992; Verheijen 1997).  
Thus we should expect some increase in \A\ simply
from the change of circular velocity measures.

Indeed, it is reassuring that the relation
found here is so closely consistent with the previous version.
McGaugh \etal\ (2000) chose \ML\  by a different method, taking a constant
value for all galaxies motivated by population models.  More importantly,
the data are largely independent:
the BTF of McGaugh \etal\ (2000) included 110 galaxies from 
Bothun \etal\ (1985), 14 from Matthews, van Driel, \& Gallagher (1998), 
and 65 from Eder \& Schombert (2000) that are completely distinct from 
and independent of the data in Table 1.  The agreement of such diverse and
independent data sets provides strong confirmation of the basic empirical result.

\subsection{Correlations Among Evolutionary Parameters}

The BTF is by far the strongest correlation present in the data, but it is not 
the only one.  There are a few others worth noting.  Some of these are shown
in Fig.~5.  Many others are not shown to avoid redundancy: anything
that is correlated with disk mass is also correlated with \Vf\ through the BTF.
Correlation coefficients for the quantities appearing in Fig.~5 are given in Table 3.

\begin{figure*}
\epsscale{0.8}
\plotone{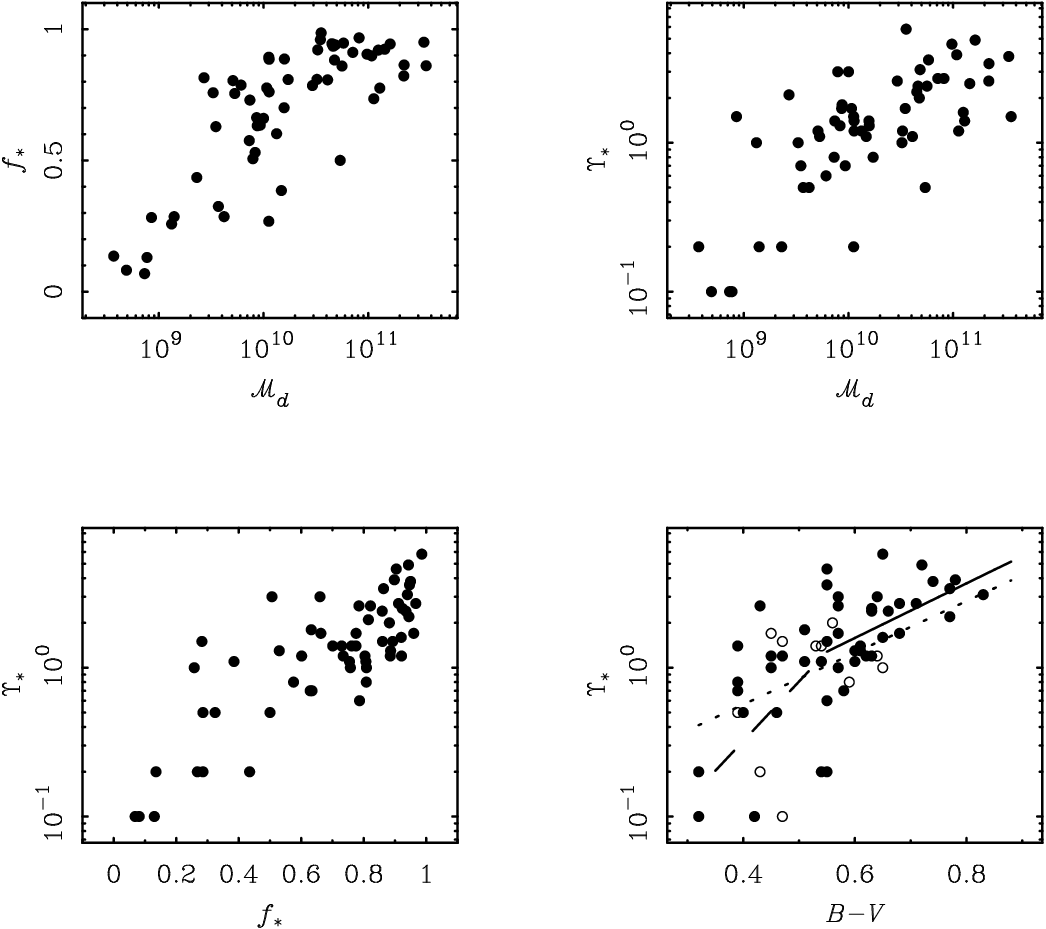}
\caption{Some of the stronger correlations in the current sample
beyond the BTF.  The stellar mass fraction $\fst$ is correlated with
disk mass and \ML\ (left panels).  The mass-to-light ratio is also 
correlated with disk mass and color (right panels).  
In the \ML---$(B-V)$ diagram at bottom right, a fit to the data
is shown as a solid line.  The fit shown is to data with
$B-V > 0.55$, where one expects a break in the \ML---$(B-V)$
relation (Portinari \etal\ 2004).  The dashed line has the slope 
expected for $B-V < 0.55$ in their models, with normalization
chosen to match the fit to the data at $B-V = 0.55$.  For comparison,
the dotted line shows the model mass-to-light ratio---color relation 
of Bell \etal\ (2003) which has been used to define $\Pop = 1$.
\label{fsMdMLBV}}
\end{figure*}

One interesting quantity that can be constructed from the data in Table 1 
is the fraction of the total observed baryonic mass in the form of stars:
$\fst = \Mst/\Md$.  The variation of \fst\ with disk mass basically shows 
the turn-away of the stellar mass TF from the BTF as one goes from star to 
gas domination.  There is lots of scatter in the \fst-\Md\ diagram
which is not reflected in the BTF.
This reiterates that what matters is mass, not the form it is in.

\begin{deluxetable}{lcccc}
\tablewidth{0pt}
\tablecaption{Correlation Coefficients\label{corrmatrix}}
\tablehead{
 & \colhead{\Md} & \colhead{\fst} & \colhead{\ML} & \colhead{$B-V$} }
     \startdata
\Vf	&	0.99 &		0.78 & 	0.77 & 0.53 \\
\Md &	\nodata &	0.76 & 	0.76 & 0.53 \\
\fst	&	\nodata & \nodata &	0.84 &	0.54 \\
\ML & \nodata & \nodata & \nodata & 0.64 \\
	\enddata
\end{deluxetable}

Together with the mass-to-light ratio and color, \fst\ is an
indicator of the evolutionary state of a galaxy (McGaugh \& de Blok 1997, 1998a;
Schombert, McGaugh, \& Eder 2001).  As evolution proceeds and a galaxy
converts its initial gas into stars, the stellar fraction increases.  As old stars
accumulate, the mean color reddens, and the mass-to-light ratio increases.
One therefor expects these quantities to be related.

The evolutionary quantities \fst, \ML, and color do indeed correlate as expected.
Not only does the mass-to-light ratio increase as the stellar population reddens,
so too does \ML\ climb as \fst\ increases.  Indeed, the \ML-\fst\ correlation is the
next strongest in Table 3 after the BTF.  

In this sample, the evolutionary quantities are correlated with baryonic mass.
They are also correlated with \Vf, as shown previously by Sanders (1996).
This must follow, given the BTF.  Indeed, figures constructed with \Vf\ instead
of \Md\ look so similar that they are difficult to distinguish.

One curious feature of the correlations with disk mass is that all of the evolutionary
indicators suggest that more massive disks are typically more evolved.
This is opposite what one might naively expect from a hierarchical galaxy
formation picture.  In such a scenario, small galaxies form first so should have
evolved the most.  That the opposite is true seems more consistent with monolithic 
or even top-down galaxy formation.  Similar results have been found recently
at high redshift (e.g., Juneau \etal\ 2005; Treu \etal\ 2005), where the tendency
for massive galaxies to be the most evolved is called ``downsizing.''
If galaxies do form hierarchically,
the observed trend suggests that the mechanisms which regulate a galaxy's
post-formation evolution dominate over the formation epoch in determining
its present evolutionary status.  This would appear to be a continuous function of
mass, given the continuity in Fig.~5.  It does not appear to be as simple as a single 
epoch of reionization suppressing galaxy formation at a characteristic mass scale.

One word of caution is that while these data span a broad range of galaxy 
properties, they do not constitute a complete volume limited sample.  
The details of these correlations may well change as other data are added.  
Nevertheless, it seems unlikely that the trend apparent in this figure could 
be reversed:  massive galaxies still in the early stages of evolution seem to be
very rare.

It is extremely difficult to obtain a sample extending to very low mass which is
complete in any meaningful sense.  Since an \HI\ map is a prerequisite for
membership in this sample, it seems likely that there are some low mass, 
high \fst\ galaxies which are not represented here.  So while there are some 
clear correlations with disk mass \textit{in this sample}, there is no guarantee 
that precisely these correlations will hold
for all galaxies.  There is more hope that correlations between the evolutionary 
parameters \fst, \ML, and $B-V$ will hold, if only because they make so much sense.

These cautions do not apply to the BTF.  
There is no hint of deviation from it: the residuals are small and uncorrelated
with any measured parameter.  There is no reason to suspect that any rotating
galaxy deviates from the BTF --- even those that should deviate do not 
(McGaugh \& de Blok 1998a; Verheijen 2001).  
The BTF appears to be a fundamental relation.

\subsection{Consistency with Stellar Population Models}

The remarkable consistency of the \MDAC\ mass-to-light ratios with
stellar population synthesis models has been noted previously 
(Sanders 1996; McGaugh \& de Blok 1998b; Sanders \& McGaugh 2002;
McGaugh 2004).  I have added more color information to the data in Table 1 than
has previously been available, and this point remains true.  Indeed, there are now
enough data that we can perform a fit in the same manner as done for
population synthesis models.  Following the format of Bell \& de Jong (2001), 
the data are fit to a relation of the form
\begin{equation}
\log \ML = a - b(B-V).
\end{equation}
Portinari \etal\ (2004) note that while their $B$-band model \ML\ are linearly
correlated for $B-V > 0.55$, they show a break at this point. \ML\ turns down to
lower values for bluer colors than predicted by extrapolation of the line fit to
redder colors.  I therefor consider two fits to the data: one covering data of all
colors, and the other restricted to $B-V > 0.55$  

The relations from the stellar population synthesis models of Bell \etal\ (2003) 
and Portinari \etal\ (2004) are compared with the fits to the data in Table 4.
The relation of Bell \etal\ (2003) is that used to estimate \MLpop\ (equation 7).
Note the close agreement between the two models: these are virtually
indistinguishable in Fig.~5.  The zero points $a$ of the models are dominated 
by the choice of IMF, which is fortuitously close (Kroupa \& Weidner 2003).

\begin{deluxetable}{lccl}
\tablewidth{0pt}
\tablecaption{Mass-to-Light Ratio--Color Relations\label{MLrlns}}
\tablehead{
  \colhead{Source} & \colhead{$a$} & \colhead{$b$} & \colhead{Note} }
     \startdata
Bell \etal\ & $-0.942$ & 1.737 & Scaled Salpeter IMF \\
Portinari \etal\ & $-0.925$ & 1.69\phn & Kroupa IMF \\
Fit to the Data & $-0.90\phn$ & 1.83\phn & $B-V > 0.55$ \\
Fit to the Data & $-1.22\phn$ & 2.31\phn & All colors \\
	\enddata
  \tablecomments{Relations of the form \\ $\log\ML = a+b (B-V)$.}
\end{deluxetable}

There is close agreement between the parameters $a$ and $b$ fit to the data
and those predicted by the models.
The \MDAC\ mass-to-light ratios, determined by dynamical methods completely
independent of the population models, show precisely the same trends.  A slight
offset in normalization is apparent, but its formal significance is low.  It is tempting
to interpret this offset as the molecular gas which has been subsumed into \MLopt.
If the mass of gas in the molecular phase is typically 10\% to 20\% of that
in stars, the offset between models and data would be reconciled.  

The break at $B-V \approx 0.55$ in the models of Portinari \etal\ (2004) 
is apparent in the data.  Fig.~5 shows (as the dashed line)
the model slope for bluer colors (from their Fig.~28), 
normalized to match the fit to the data at the break point.
This is entirely consistent with the downward trend in the data.

The scatter in \ML\ is also as expected, being larger in $B$ than in $K'$
(Sanders \& McGaugh 2002; McGaugh 2004).  This is true also above and
below the break point at $B-V = 0.55$.  Below this point, one expects a
tremendous amount of scatter from the rapid evolution of young populations.
Here the scatter is enormous:  $\sigma_{\ML} = 0.52$.  Above this color,
one expects variation in \ML\ to settle down, as the effects of individual star
formation events are moderated by the accumulation of mass
from previous generations.  For these redder colors, the scatter is
$\sigma_{\ML} = 0.20$.

In sum, the $\Q = 1$ mass-to-light ratios are optimal not only in terms of 
minimizing the scatter in the BTF and the \MDAC\ (from which
they come), but also in terms of our expectations for stellar populations. Indeed,
it is hard to imagine better agreement with independent population models to
which no fit has been made.  Moreover, the scatter in the dynamical relations
is so small for these high quality data that \ML\ estimated from the BTF itself
(equation 7) are nearly indistinguishable from \MLopt\ from the \MDAC.
Either empirical method can be employed with unprecedented precision.

\subsection{Test by Extrapolation} 

The value of the slope $x$ of the BTF is somewhat controversial, 
being of considerable physical importance.
The nominal expectation of CDM is that $x = 3$ (Navarro \& Steinmetz 2000a,b),
and Courteau \etal\ (2003) argue that this is consistent with their $I$-band data.  
The $I$-band is a good but not perfect indicator of mass, and the HST calibrated
distances of Sakai \etal\ (2000) give an $I$-band slope of 4.
This emphasizes the need for a good estimate of \Md\ and not just $L$.

McGaugh \etal\ (2000) found $x = 4$ from data and \ML\ estimates independent
of those used here.  The key aspect of that study was the large number of
very gas rich, low rotation velocity galaxies which tied down the low mass end
of the BTF.  Verheijen (2001) obtained the same result.
Bell \& de Jong (2001) generated stellar population models in an attempt to
improve the estimate of \ML.
Applied to the data of Verheijen, they found a BTF with $x = 3.5$.
I have, in effect, repeated this analysis with more data and updated models, 
and find very much the same result: for $\Pop = 1$, $x = 3.4$ (Table 2).  

It is interesting that while the models and data are in good agreement (\S 4.2), 
use of \MLpop\ gives somewhat shallower $x$ than \MLopt\ or \MLmax\
(Fig.~3).  This is attributable in part to the slight difference in normalization
($a$ in Table 4).  This can not be the entire reason for the difference, as
increasing \Pop\ to 2 only increases $x$ to 3.7.  The shortcoming of
\MLpop\ estimated from a single color is that galaxies become beads on a string
in the \ML-$(B-V)$ diagram: all the data collapse to fall on the dotted line in Fig~5.  
The inability to estimate deviations from the mean relation and thereby include
a realistic estimate of the scatter seems to result in a bias\footnote{One could,
perhaps, do better with more colors.  The incorporation of such information had
an analogous effect on the slope of the $K'$-band \ML-color relation, which 
became flatter between Bell \& de Jong (2001) and Bell \etal\ (2003).}
in the  determination of the slope.  Relative to the population 
predicted \MLpop, \MLopt\ is skewed to high values for red colors and to low 
values for blue colors.  One expects such a skew relative to a single straight
line fit (Portinari \etal\ 2004); it causes a systematic difference in the slope in spite 
of the close agreement seen in Fig.~5 and Table 4.

The critical issue for constraining the slope is the dynamic range of the data.  
As a further test of the slope of the BTF, I have sought out galaxies that extend
the relation to lower rotation velocities.  The slowest rotator in Table 1 has
$\Vf = 54\, \kms$.  I have searched the literature for galaxies with slower
rotation velocities that are of adequate quality to make a useful comparison.
There are many slow rotators already in the sample of McGaugh \etal\ (2000),
but there the rotation velocity estimate was based on a line-width (Eder \& 
Schombert 2000).  Here I require that there be a resolved measurement
adequate for estimating \Vf.

Table 5 contains the objects found that met these criteria.  
Many of these are due to the recent excellent work of Begum \& 
Changular (2003, 2004a,b) which extends down to objects approaching
the globular cluster mass scale.  In Table 5 I give the data necessary for
the BTF from the data given in each reference cited there.  
I have estimated \Vf\ myself from the published data.
In addition to the best estimate, I give a generous range of uncertainty.
In many of these cases, there is a substantial correction for asymptotic drift.
The lower limit of \Vf\ takes the lower end of the published errors ignoring the
asymptotic drift correction.  The upper limit takes the upper end of the experimental
errors and includes the correction.  This procedure results in a very broad error
estimate:  the uncertainties listed in Table 5 are much larger than $1 \sigma$.

\begin{deluxetable}{lcccc}
\tablewidth{0pt}
\tablecaption{Extreme Dwarf Galaxy Data\label{BCdata}}
\tablehead{
\colhead{Galaxy} & \colhead{\Vf} & \colhead{\Mst} & \colhead{\Mg} &
 \colhead{Ref.} \\
   \colhead{} & \colhead{(\kms)} & \multicolumn{2}{c}{($10^{6}\;\Msun$)} &
   \colhead{} }
     \startdata
ESO215--G?009 & $51^{\phn +8}_{\phn -9}$
 & \phn 23\phd\phn    &  714\phd\phn & \phn\phd 1 \\
UGC 11583\tablenotemark{a}  & $48^{\phn +3}_{\phn -4}$
 & 119\phd\phn  & \phn 36\phd\phn  & 2,3 \\
NGC 3741 & $44^{\phn +4}_{\phn -2}$
 & \phn 25\phd\phn & 224\phd\phn & \phn\phd 4 \\
WLM	& $38^{\phn +5}_{\phn -5}$
	& \phn 31\phd\phn	& \phn 65\phd\phn	& \phn\phd 5 \\
KK98 251  & $36^{\phn +8}_{\phn -4}$
    & \phn 12\phd\phn       & \phn 98\phd\phn    & \phn\phd 3 \\
GR 8    & $25^{\phn +5}_{\phn -4}$
    & \phn\phn 5\phd\phn    & \phn 14\phd\phn    & \phn\phd 6 \\
Cam B   & $20^{+10}_{-13}$
    & \phn\phn 3.5          & \phn\phn 6.6      & \phn\phd 7 \\
DDO 210 & $17^{\phn +3}_{\phn -5}$
    & \phn\phn 0.9          & \phn\phn 3.6      & \phn\phd 8 \\
\enddata
\tablenotetext{a}{UGC 11583 = KK98 250.}
\tablerefs{1.~Warren, Jerjen, \& Koribalski (2004).
2.~McGaugh, Rubin, \& de Blok (2001).
3.~Begum \& Chengalur (2004a).
4.~Begum, Chengalur, \& Karachentsev (2005).
5.~Jackson \etal\ (2004).
6.~Begum \& Chengalur (2003).
7.~Begum, Chengalur, \& Hopp (2003).
8.~Begum, \& Chengalur (2004b). }
\end{deluxetable}

For the stellar and gas mass, I take the value given by the original authors.
\Mst\ has often been estimated using the models of Bell \& de Jong (2001),
so these correspond roughly to $\Pop =1$ estimates.  Rather than an uncertainty,
I consider the full range of possible stellar masses, from zero to maximum disk.
These are the vertical lines in Fig.~6.

The extrapolation of the BTF fit to the data in Table 1 does an excellent job
of predicting the data for these lower mass galaxies.  The objects in Table 5
follow the slope $x = 4$ down to unprecedented low velocity.  
The BTF remains valid over five decades in mass.

\begin{figure*}
\epsscale{0.8}
\plotone{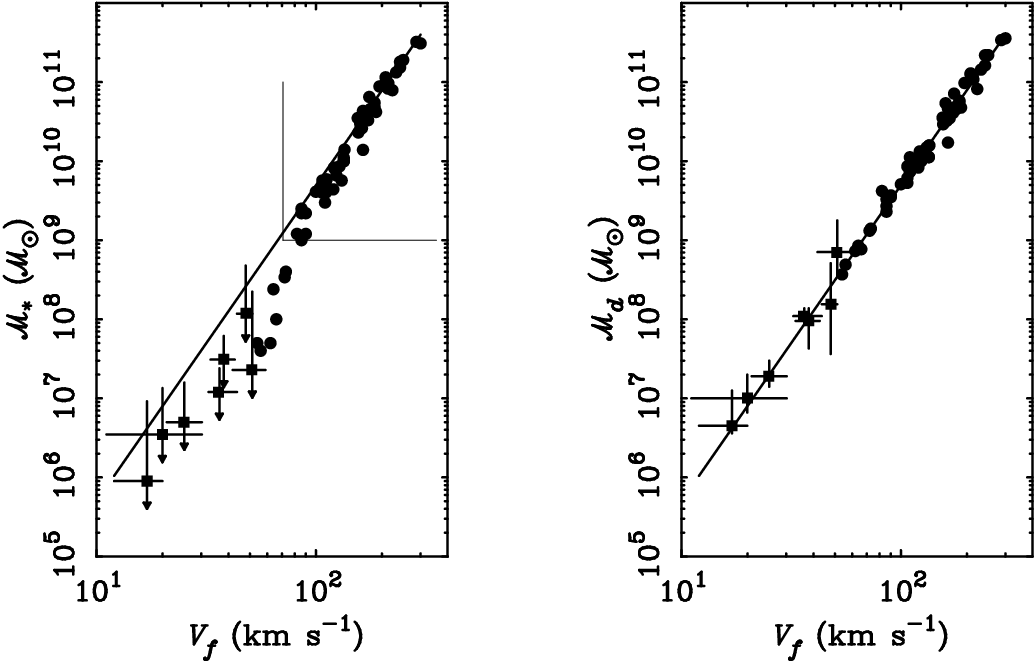}
\caption{The stellar mass (left) and baryonic (right) Tully-Fisher relations, 
including the data for the extreme dwarf galaxies listed in Table 5.
The horizontal lines through these objects are the maximum plausible range 
for \Vf:  these are much larger than $1 \sigma$ error bars.  The vertical lines 
show the full range of possible stellar masses, from zero to maximum disk.
The extrapolation of the BTF fit to the more massive galaxies from Table 1 
is in good agreement with these extreme dwarfs.  The importance of this
check is illustrated by the thin lines inset in the left panel. These show the
limits of samples that suggest shallower slopes (e.g.,
Bell \& de Jong 2001; Courteau \etal\ 2003).
\label{BTFlong}}
\end{figure*}

Fig.~6 also illustrates why the slope has been difficult to constrain.
The lower limit of other studies ($\sim 10^9\,\Msun$, $\Vf \approx 70\,\kms$:
e.g., Bell \& de Jong 2001; Courteau \etal\ 2003)
is shown in the left hand panel.  The data considered there extend over only
a small fraction of the range studied here (down to $5 \times 10^6\,\Msun$
with the extreme dwarfs in Table 5).  This lack of dynamic range probably
dominates all other factors (such as the mass determination method or the
velocity estimator used) in constraining the slope.  Indeed, if we were to
truncate the data in Table 1 at the same level, we would fail to perceive the
break-point in the stellar mass Tully-Fisher relation.  Such a sample
would be dominated by star-dominated galaxies, and fail to provide the
constraint on \ML\ which follows when gas-dominated objects are included.

The steep slope of the BTF should come as no surprise,
as it is completely consistent with the results of 
McGaugh \etal\ (2000).  That study made use many low mass galaxies,
19 of which have $\onehalf W_{20} < 70\,\kms$. 
Those objects are completely independent of the galaxies discussed here.  
The new, more accurate data in Table 5 simply return the same result with 
less scatter.  Other workers investigating low mass galaxies
have also inferred the need for a steep slope
(e.g., Gurovich \etal\ 2004; Pizagno \etal\ 2004).

Formally, a slope as shallow as $x = 3$ deviates from the optimal BTF by
$7 \sigma$.  This can be made less by changing \ML, but only at the price
of degrading the correlation and the the many consistency checks on \ML.  
In order to recover a slope as shallow as $x = 3$, one requires $\Pop = 0.36$ or
$\G < 0.1$ (Table 2).  Such absurdly sub-maximal disks would fall considerably
short of the mass which is directly observed in stars locally.  

Consideration of the extreme dwarfs renders it even more difficult to reconcile
a shallow slope with the data.  While it is possible, at least in principle, to move 
massive galaxies down in mass by reducing their mass-to-light ratios in order
to accomodate a shallow slope, it is not possible to move the extreme dwarfs
very far up.  Even taking maximum disk in those cases makes little difference
to the slope, and causes the curious situation that the IMF in these low mass
galaxies must be systematically heavier than that in giant galaxies.
It is thus very difficult to reconcile a shallow ($x = 3$) slope for the BTF 
with the data.

\subsection{The Maximality of Disks}

One application of the result here is to quantify the degree to which galaxy
disks are maximal.  There is considerable debate as to whether high surface
brightness disks are maximal (e.g., Sellwood 1999; Courteau \& Rix 1999).
There would seem little doubt that low surface brightness disks are dark matter
dominated (de Blok \& McGaugh 1997, 2001), but an argument for maximal 
disks can be made even in these objects (Fuchs 2003).
It is therefor of considerable interest to investigate how maximal disks are,
and how disk maximality varies with disk properties.

\begin{figure*}
\epsscale{1.0}
\plotone{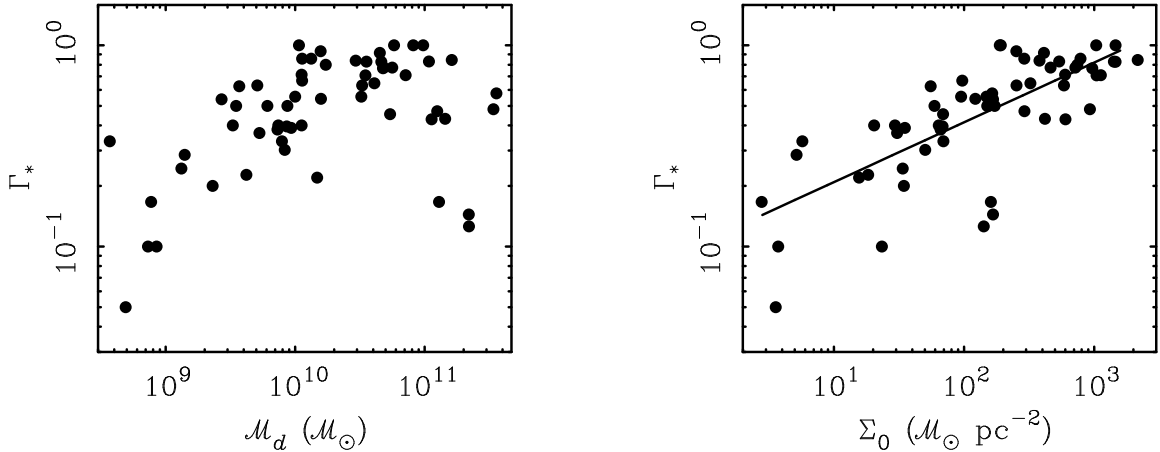}
\caption{The maximality of disks (\Gopt\ for $\Q =1$) as a function of disk mass 
(left) and the central stellar surface mass density (right).  
While the correlation with mass
is poor, there is a clear correlation of maximality with disk surface density. 
High surface density disks tend to be nearly maximal, 
while lower surface density disks are systematically sub-maximal.  
The line is a fit to the data (equation 12).
\label{Gopt}}
\end{figure*}

For the \MDAC\ mass-to-light ratios favored here, 
the fraction of maximum disk in each case is
\begin{equation}
\Gopt = \frac{\MLopt}{\MLmax}.
\end{equation}
This is plotted against disk mass and surface density in Fig.~7.
There is only a weak correlation of \Gopt\ with disk mass (${\cal R} = 0.45$)
which depends heavily on rather few points at low mass (cf.\ Persic \& Salucci 1988).  
Dynamical arguments stemming from the adherence of low surface brightness 
galaxies to the Tully-Fisher relation (Zwaan \etal\ 1995;
Sprayberry \etal\ 1995; Hoffman \etal\ 1996) suggest that disk 
maximality \Gopt\ should correlate with surface brightness 
(Tully \& Verheijen 1997; McGaugh \& de Blok 1998a; Zavala \etal\ 2003).  
We can improve on this by using the mass-to-light ratios \MLopt\ to convert the
observed central surface brightness into the central surface mass density
of stars:
\begin{equation}
\log \Sd = \log \MLopt +0.4 (27.05-\csb).
\end{equation}

As anticipated, there is a good correlation between \Gopt\ and \Sd\ 
(${\cal R} = 0.74$).  A fit to the data in Fig.~7 gives
\begin{equation}
\log\Gopt = -0.98+0.3 \log\Sd.
\end{equation}
In terms of the more directly observable central surface brightness,
this translates to $\log\Gopt = 3.13 - 0.16 \csb$.
There is considerably greater scatter about this latter relation.

Irrespective of how we frame the relation, or what mass-to-light ratio
prescription we prefer, it seems inevitable that 
the disk contribution must decline systematically as surface density declines.
Low surface brightness disks are inevitably dark matter dominated.  
In contrast, high surface density disks contribute a non-negligible 
fraction of the total mass at small radii for plausible \ML.  
Remarkably, this leaves no residual signature in the Tully-Fisher relation
(McGaugh \& de Blok 1998a; Courteau \& Rix 1999) in spite of the generally
modest radius at which rotation curves achieve \Vf.

Statistics of these data, divided into quartiles by \Sd, are given in Table 6.
The typical $\Gopt = 0.78$ in the highest surface density quartile.  
\Gopt\ can not exceed unity, and is projected to saturate at 
$\csb \approx 19.5\,\magsq$.  This is comparable to the highest surface 
brightness disks that exist (Marshall 2004).
In the lowest surface density quartile, the typical fraction of maximum
disk drops to $\Gopt = 0.25$.  This confirms and quantifies the well-known
result that low surface brightness disks are sub-maximal (de Blok \& McGaugh
1997).  There is no empirical indication that we have reached a lower limit
in \Gopt.  Disks lower in surface brightness than the most extreme
considered here do exist, though they have yet to be observed with 
sufficient accuracy to be included here.  

\begin{deluxetable}{ccccc}
\tablewidth{0pt}
\tablecaption{Maximality of Disks\label{maximality}}
\tablehead{
\colhead{Quartile} & \colhead{$N$} & \colhead{$\langle \csb^B \rangle$} &
  \colhead{$\langle \Sd \rangle$} & \colhead{$\langle \Gopt \rangle$} \\
  \colhead{} & \colhead{} & \colhead{(\magsq)} &
   \colhead{(\surfdens)} & \colhead{}  }
   \startdata
1	& 15	& 23.31	& \phn 21	& 0.25 \\
2	& 15	& 22.31	& \phn 99	& 0.48 \\
3	& 15	& 21.52	& 295	& 0.74 \\
4	& 15	& 20.41	& 964	& 0.78 \\
\enddata
\tablecomments{The biweight location of the 
the disk central surface brightness, 
central stellar mass surface density, 
and the degree of maximality of disks
are given for each quartile of the sample.}
\end{deluxetable}

The meaning of \Gopt\ for high surface brightness galaxies is subject to the
caveats discussed in \S 3.1.  In particular, the effective definition of maximum 
disk typically accounts for 84\% rather than 100\% of the velocity at the peak
of the disk contribution (Sackett 1997).  If the disks in the highest quartile
have 78\% of this mass, then they contribute 74\% of the velocity at 
2.2 scale lengths ($\mass \propto V^2$).  
This is not very different from the sub-maximal
contribution of 63\% advocated by Bottema (1993) and is certainly within
the galaxy-to-galaxy scatter.  Kregel, van der Kruit \& Freeman (2005) 
find a slightly lower
mean velocity contribution, but their sample is dominated by intermediate
surface brightness galaxies, so such a result is consistent with the trend
apparent in Fig.~7.  

The results in the literature seem broadly consistent, bearing in mind that
a good deal depends on what is really meant by ``maximum disk.''  The
most important point here is that the degree of maximality of a disk depends
systematically upon its surface density.  There is no magic value of \Gopt\
that is a fixed fraction for all disks (Bottema 1997; de Blok \& McGaugh 1996).

\section{Conclusions}

I have explored the Baryonic Tully-Fisher relation for many choices of stellar
mass-to-light ratios using a sample of high quality data spanning a large
dynamic range in mass.  I provide fits to the BTF for each \ML.  There is
a particular choice, based on the minimization of the scatter in the local
mass-discrepancy---acceleration relation (the \MDAC: McGaugh 2004), 
that also minimizes the scatter in the BTF.  This optimal BTF is
\begin{displaymath}
\Md = 50 \Vf^4
\end{displaymath}
with \Vf\ in \kms\ and the total baryonic mass of a galaxy in \Msun.
This provides a remarkably precise method of estimating the mass of
rotating galaxies by observation of a single global observable, the
level at which the rotation curve becomes flat.

The form in which the baryonic mass resides, stars or gas, makes no difference
to the BTF.  Only the sum matters.  This strongly suggests that the BTF is a
fundamental relation between rotation velocity and baryonic mass.
It further implies that there is
no other large reservoir of baryons which matter to the sum:  the
stars and gas observed in spiral galaxies account for essentially all of the
baryonic mass therein.  

The mass-to-light ratios determined for the optimal BTF are in exceptionally
good agreement with stellar population synthesis models.  This consistency,
together with the agreement between local and global empirical relations
connecting baryonic mass to the observed dynamics, implies that the baryonic
mass is well determined.  This would appear to solve the long standing problem
of the uncertainty in the mass of stellar disks.

Using these robust stellar mass estimates, I have examined a variety of
evolutionary measures.  The stellar fraction, mass-to-light ratio, and color
all correlate with each other as one would expect.  Little evolved galaxies
with low \fst\ tend to have blue colors and low \ML; more evolved galaxies
have higher stellar fractions, redder colors, and higher mass-to-light ratios.
These quantities are also correlated with disk mass and rotation velocity:
more massive disks tend to be more evolved.

The degree to which disks are maximal 
varies systematically with stellar surface density.  
High surface brightness galaxies tend to be more nearly maximal, typically with
$\ML \sim 78\%$ of the maximum disk value at $\csb^B = 20.4\; \magsq$. 
Low surface brightness galaxies are sub-maximal, with
$\ML \sim 25\%$ of maximum disk at $\csb^B = 23.3\; \magsq$.
There is considerable variation from galaxy to galaxy.
In no case is the stellar mass completely negligible at small radii,
a fact that is important to mass models and
constraints on the inner slope of the halo mass distribution (core or cusp).

\acknowledgements The work of SSM is supported in part by NSF grant 
AST0206078 and NASA grant NAG513108.  
SSM is grateful for conversations on this subject with many people,
especially Greg Bothun, Bob Sanders, Brent Tully, and Rob Swaters.
SSM thanks the Astronomy department of Case Western Reserve University 
for its hospitality during a sabbatical visit when much of this work was done.

%\clearpage

\end{document}